# Comparison between InAs-based and GaSb-based Interband cascade lasers with hybrid superlattice plasmon-enhanced claddings

B. Petrović,[1] A. Bader,[1] J. Nauschütz,[2] T. Sato,[3] S. Birner,[3] S. Estevam,[1] R. Weih,[2] F. Hartmann[1] and S. Höfling[1]

_borislav.petrovic@physik.uni-wuerzburg.de_       _fabian.hartmann@physik.uni-wuerzburg.de_

[1]*Julius-Maximilians-Universität Würzburg, Physikalisches Institut, Lehrstuhl für Technische Physik, Am Hubland, 97074 Würzburg, Germany*

[2]*nanoplus Advanced Photonics Gerbrunn GmbH, Oberer Kirschberg 4, 97218 Gerbrunn, Germany*

[3]*nextnano GmbH, Konrad-Zuse-Platz 8, 81829 München, Germany*

We compare InAs-based and GaSb-based interband cascade lasers (ICLs) with the same 12 stages active region designed to emit at a wavelength of 4.6 µm. They employ a hybrid cladding architecture with the same geometry and inner claddings consisting of InAs/AlSb superlattices but different outer claddings: The InAs-based ICL employs plasmon enhanced n-type doped InAs layers while the GaSb-based ICL employs plasmon-enhanced n-type doped $InAs_{0.915}Sb_{0.085}$ claddings lattice matched to GaSb. Due to the lower refractive index of n⁺-InAsSb ($n = 2.88$) compared to n⁺-InAs ($n = 3.10$), and higher refractive index of separate confinement layers[29], the GaSb-based ICL shows a 3.8 % higher optical mode confinement in the active region compared to the InAs-based ICL. Experimentally, the GaSb-based ICL shows a 17.3 % lower threshold current density in pulsed operation at room temperature. Also presented is the influence of geometry and doping variation on confinement factors and calculated free carrier absorption losses in the GaSb-based ICL.

## I. INTRODUCTION

Interband cascade lasers (ICLs[1]) have emerged as prominent coherent mid-infrared light sources for various applications such as industrial process control, breath analysis, IR scene projection and military applications[2-8]. The broken band alignment of wells in their recombination region makes the emission wavelength easily tunable in the mid-infrared range by changing the widths of the active wells. Recycling electrons through cascaded stages for multiple optical transitions, firstly introduced for quantum cascade lasers (QCLs), in combination with long upper-level lifetime of interband transitions provides them with low threshold current densities in the 3-6 µm wavelength range[8-9]. In state-of-the-art ICLs, InAs/AlSb superlattice (SL) claddings are commonly used due to their low refractive index ($n = 3.39$[10]) to confine the light to the active region and their possibility to be grown strain compensated on GaSb and InAs substrates. An important milestone in cladding design was the introduction of lightly doped separate confinement layers and moderate, gradual doping of claddings in 2008[11], when the first continuous wave (cw) operation at room temperature was demonstrated at a wavelength of 3.75 µm. Towards longer wavelengths, a proper design of the cladding layers becomes more demanding as the free carrier absorption (FCA) losses increase with wavelength squared. Heavily doping of InAs emerged as beneficial for a significant reduction of refractive index, making it a good candidate for a cladding material[12-13]. Use of plasmon-enhanced InAs in combination with inner InAs/AlSb claddings in IC lasers enabled lasing emission at wavelengths as long as 11.1 µm in 2015[14] (at 130 K), further extended to 13.2 µm in 2022[15] (at 115 K) and 14.4 µm in 2023[16] (at 120 K) by introduction of an additional pair of InAsP barriers in the active region. The same concept of hybrid claddings containing either plasmon-enhanced InAs or InAsSb layers has been applied to ICLs with shorter wavelengths (3.5-7 µm) to enhance their performance[17-19]. The lowest obtained threshold current densities of the ICLs of such hybrid cladding design at respective wavelengths are 247 A/cm² (at 4.6 µm)[20] and 138 A/cm² (at 3.8 µm)[21].

## II. THEORETICAL ANALYSIS

For a hybrid cladding architecture, we are investigating the benefits of ICLs grown on GaSb substrates in comparison to ICLs grown on InAs substrates. Outer, plasmon-enhanced claddings for InAs-based and GaSb-based ICL are comprised of highly doped InAs and $InAs_{0.915}Sb_{0.085}$ respectively. Figure 1 (a) shows the energy dispersion of the first states in bulk InAs and $InAs_{0.915}Sb_{0.085}$ calculated by the $\mathbf{k \cdot p}$ model in the nextnano software[22]. The presence of Sb in the alloy reduces the bandgap, hence $InAs_{0.915}Sb_{0.085}$ becomes transparent at wavelengths $\lambda \geq 4.3\ \mu m$ whereas pure InAs is transparent for a slightly wider wavelength range, $\lambda \geq 3.5\ \mu m$. In this paper we will focus on emission at $\lambda = 4.6\ \mu m$, as the wavelength is in the atmospheric transparency window and also matches the spectral line of carbon-oxide (CO). Based on this, refractive index dispersion of InAs and InAsSb can also be derived from the Lorentz-Drude model[23]. Figure 1 (b) shows the refractive indices for different emission wavelengths and different doping levels for InAs and InAsSb. InAsSb shows a shorter plasmon cutoff wavelength compared to InAs for identical doping concentrations. This is a result of a lower effective mass in InAsSb compared to InAs[23,24]. Consequently, at a certain wavelength $InAs_{0.915}Sb_{0.085}$ provides a smaller refractive index than InAs for the same charge carrier density.

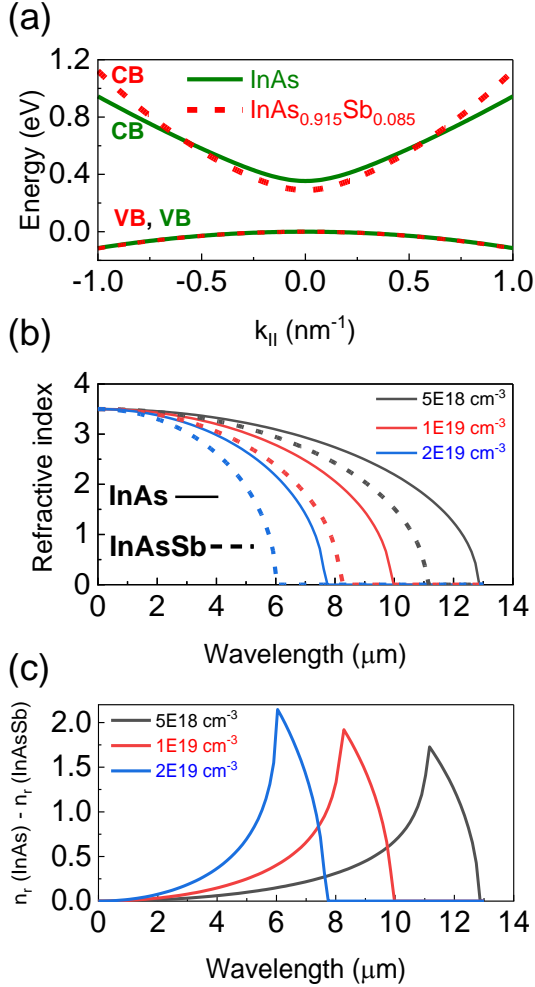

FIG. 1. (a) Band dispersion of InAs and $InAs_{0.915}Sb_{0.085}$. (b) Dispersion of refractive indices for different doping concentrations of $n^+$-InAs and $n^+$-InAsSb. (c) Differences between refractive indices of $n^+$-InAs and $n^+$-InAsSb for different doping concentrations.

In Figure 1 (c) differences between their refractive indexes are shown for different doping concentrations. Their values are positive for emission wavelengths up to the plasmon cutoff wavelength of InAs. The differences in refractive indices significantly increase with emission wavelength. The maximum difference coincides with the plasmonic wavelength of InAsSb. In order to validate the considered cladding design for GaSb-based ICL, several parameters have been investigated. Figure 2 depicts extracted refractive indices of $InAs_{0.915}Sb_{0.085}$ for different charge carrier densities at emission wavelength $\lambda = 4.6$ µm. A linear decrease in refractive index is observed with increasing carrier concentration. FCA losses can be calculated by taking into account a non-linear dependance of carrier mobility on charge carrier density using a modified Hilsum formula [C. Hilsum. Simple empirical relationship between mobility and carrier concentration. Electr. Lett. 10, 259 (1974.)] introduced for highly doped InAs by A. Baranov[23], eq (1)

(1) $$\mu_e = \frac{3 \times 10^4 \text{ cm}^2/\text{Vs}}{1+(\frac{N}{8\cdot 10^{17}})^{0.75}}.$$

Here, $\mu_e$ is the electron mobility, N is the carrier concentration. Figure 2 also shows the calculated FCA losses[23-25] as calculated in Ref. 23. An exponential increase in FCA is observed for increasing carrier density. In this paper we used carrier concentration of $N = 1 \cdot 10^{19}$ cm$^{-3}$.

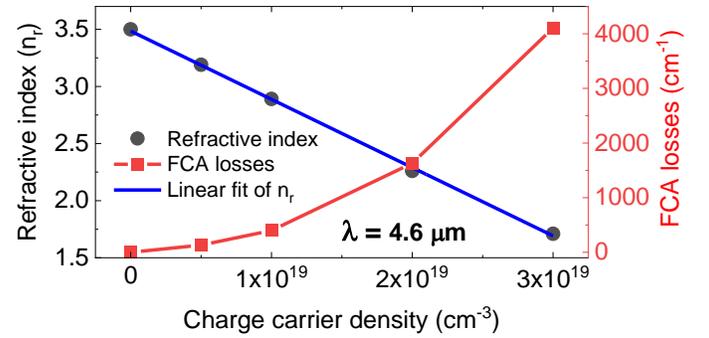

FIG. 2. Charge carrier density dependance of refractive index of $InAs_{0.915}Sb_{0.085}$ and corresponding FCA losses.

This result allows to calculate the FCA of $n^+$-InAsSb incorporated in the ICL as the outer cladding material. The role of inner InAs/AlSb superlattice claddings is to spatially separate the active region from the plasmon enhanced layers of a high optical loss. Figures 3 (a) and (b) show optical confinement factors distribution for varying thickness of the outer $n^+$-InAsSb for a fixed overall thickness of claddings. The simulation is done for thicknesses corresponding to 20/30/40/50/60/70/80 % of the total cladding thickness. For determining the optimal thickness of plasmon enhanced claddings there is a trade-off between high optical confinement and low FCA losses. For increasing the thickness of $n^+$-InAsSb claddings optical confinement in them increases exponentially, whereas for inner SL claddings it decreases parabolically. The confinement factor in the active region increases almost linearly with increasing thickness of the highly doped InAsSb layers. However, as shown in Figure 3 (c), FCA losses increase with thickness of $n^+$-InAsSb. A pronounced increase of the slope is observed for values higher than 850 nm (50 %). In the range of 340 nm (20 %) to 850 nm (50 %) there is a trade-off between higher optical mode confinement and lower FCA. An exact ideal thickness of $n^+$-InAsSb could not be identified. A difference between the edges of this interval is that a thickness of 340 nm provides 27.6 % lower FCA losses than 850 nm, but also a 4.9 % lower optical confinement factor in the active region. The optimum optical confinement factor in the active region also depends on the thickness of the separate confinement layers (SCLs).

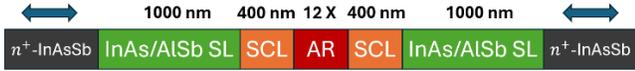

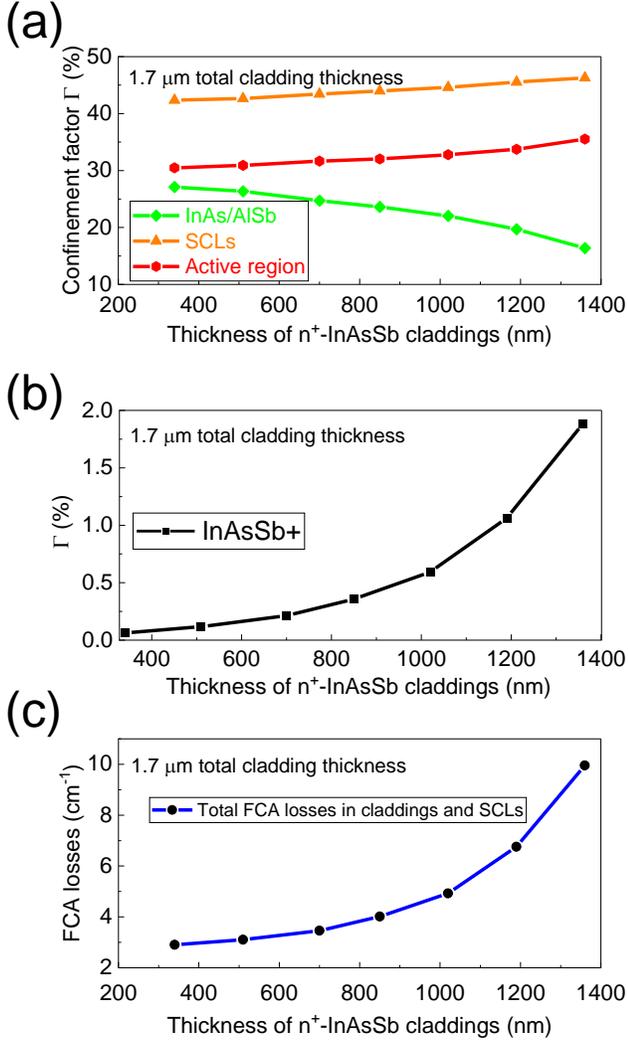

FIG. 3. (a), (b) Dependence of optical mode confinement factors in GaSb-based hybrid cladding ICL on the thickness of $n^+$-InAsSb outer claddings for 1.7 µm overall cladding thickness. In the subplot optical confinement factor in $n^+$-InAsSb claddings is shown. (c) Dependence of total FCA losses in claddings and SCLs of GaSb-based hybrid cladding ICL on the thickness of $n^+$-InAsSb outer claddings for 1.7 µm overall cladding thickness.

Figures 4 (a) and (b) depict the optical confinement factor distribution in different parts of the ICL for different thicknesses of SCLs. For increasing thickness of the SCLs from 50 to 150 nm, the optical confinement factor in the active region increases, it reaches its peak for SCL thickness of 150 nm, then for higher thicknesses it parabolically decreases. In the SCLs, the confinement increases sub-linearly, whereas in the plasmon-enhanced outer claddings and inner InAs/AlSb claddings it decreases exponentially. The total FCA losses in claddings and SCLs for different SCL thicknesses are shown in Figure 4 (c). For thicknesses up to 400 nm a strong decrease is prominent. SCL longer than 400 nm provide a slight decrease of FCA however at the expense of confinement in the active region. An SCL thickness of 400 nm provides 38 % lower FCA than for 150 nm, but 7.4 % lower active region confinement.

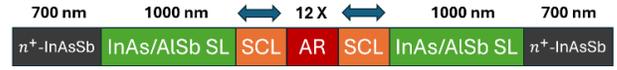

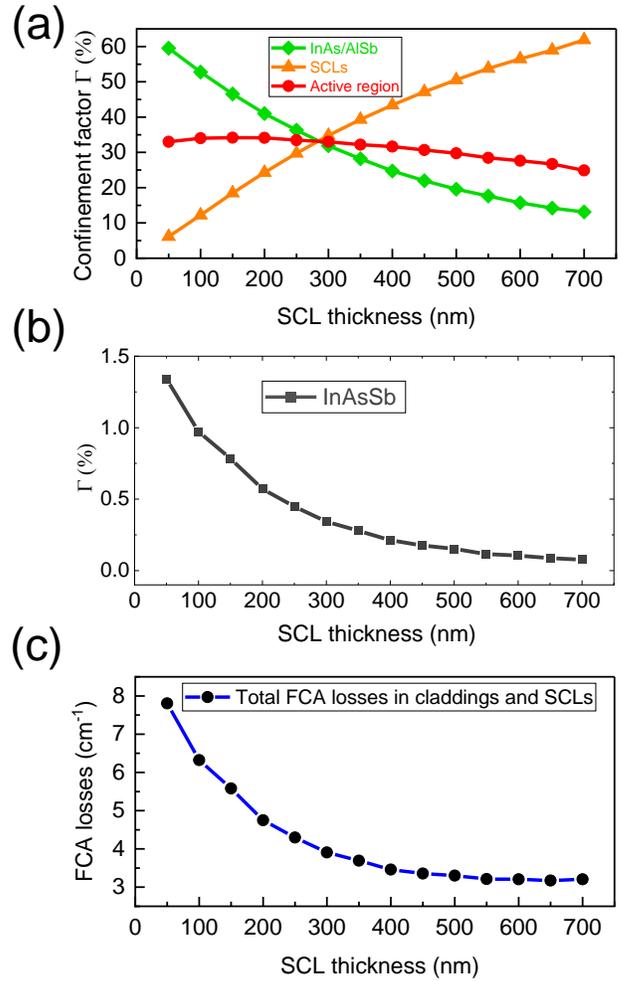

FIG. 4. (a), (b) Optical mode confinement factors in different layers of GaSb-based hybrid cladding ICL under varying thickness of SCLs. In the subplot optical confinement factor in $n^+$-InAsSb claddings is shown. (c) Total FCA losses in claddings and SCLs of GaSb-based hybrid cladding ICL under varying thickness of SCLs

However, for 400 nm thickness, the mode intensity in the vicinity of the high loss GaSb substrate is in the range of values for an ICL with state-of-the-art SL claddings while for 150 nm it is an order of magnitude higher. This suggests a bigger penetration of the optical mode into the substrate and hence it makes 400 nm thick SCLs a more promising choice. To investigate the ideal number of cascade stages, we focus on threshold power density rather than optical confinement in the active region and the FCA. ICLs lasing at 3.5 µm as the "sweet spot" of threshold current density, that have state of the art threshold power densities have open circuit voltages around 2 V[8,18,20,26,27]. From that perspective, a rule of thumb for optimal number of cascade stages in the active region is that sum of the transition energies of all the stages is approximately 2 eV. Based on that, for 4.6 µm emission, the optimal number of stages, following this criterion, would be 7. Every additional stage would ideally produce an additional photon per injected electron and hence reduce the threshold current density, however, due to increased open circuit voltage it would also increase the threshold power density. Figures 5 (a) and

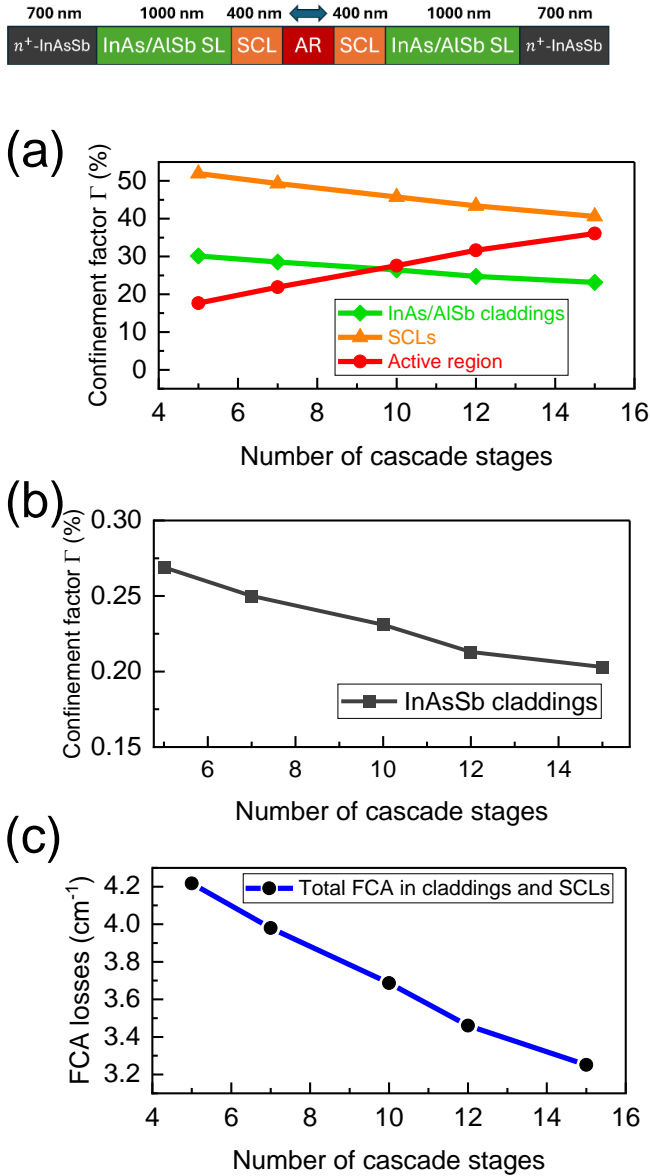

FIG. 5. (a), (b) Optical mode confinement factors in different parts of the GaSb-based hybrid cladding ICL for different number of cascade stages in the active region (c) Total FCA losses in claddings and SCLs of GaSb-based hybrid cladding ICL for different number of cascade stages in the active region.

(b) show a distribution of optical mode confinement factors in layers of the ICL for different numbers of cascade stages. By adding more cascade stages, the confinement in outer $n^+$-InAsSb claddings, inner InAs/AlSb claddings and SCLs almost linearly decreases while in the active region it increases due to widening of the recombination region. Figure 5 (c) displays the variation of the total FCA losses in claddings and SCLs with increasing the number of cascade stages. The trend is expectedly decreasing, following a decrease of confinement factors in claddings and SCLs. The 12 stages design of an ICL reported in Ref. 20 has superior mode confinement by 44.7 % and lower FCA losses by 13.1 %. compared to 7 stages. Following this trend, we conclude that design of 15 active stages would be even more beneficial for the threshold current density. However, considering that useful voltage would be higher than double in comparison to 7 stages design, it would lead to a significant increase in threshold power density. Therefore, in this paper we are considering the same 12 stage active region and cladding design as the 4.6 µm emitting ICL reported in Ref. 20. Figure 6 (a) displays the layout scheme of such ICL.

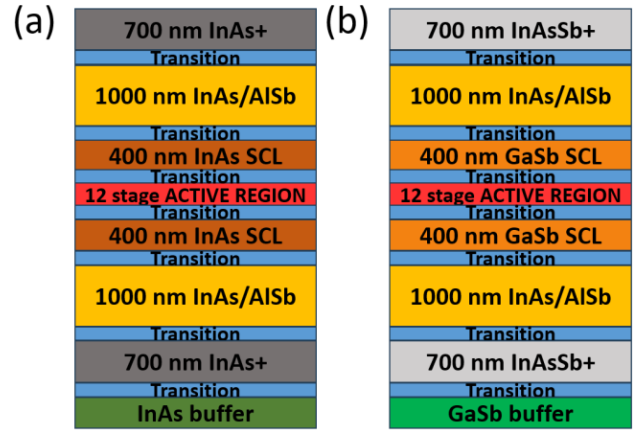

FIG. 6. Layouts of ICL structures with hybrid claddings: (a) InAs-based ICL with inner InAs/AlSb SL and outer highly doped $n^+$-InAs claddings (b) GaSb-based ICL with inner InAs/AlSb SL and outer highly doped $n^+$-InAsSb claddings.

The hybrid claddings are 1.7 µm thick and consist of 1 µm thick inner InAs/AlSb SL claddings and 0.7 µm thick outer plasmon-enhanced InAs claddings. Following the previous analysis, we conclude the same geometry can be applied to GaSb-ICL. Figure 6 (b) shows the GaSb-based ICL that has an analog architecture. The major difference from the InAs-based structure is the use of plasmon enhanced n-type doped $InAs_{0.915}Sb_{0.085}$ outer claddings lattice matched to the GaSb substrate. The GaSb-based ICL also employs GaSb separate confinement layers (SCLs) instead of InAs.

The two analog ICL structures are compared in terms of optical mode confinement in the active region and the total FCA losses. In Figure 7 (a) refractive index ($n_r$) profiles are presented contrasting InAs-based and GaSb-based

ICLs with hybrid cladding architecture[28-29]. The respective relative mode intensities (R.I.) are also displayed. The mode for GaSb-based ICL has a reduced width and hence is more strongly confined to the active region. This is attributed to the lower refractive index of highly n-type doped ($1 \cdot 10^{19} cm^{-3}$) n$^+$- InAsSb of $n_r = 2.88$, than of equally doped n$^+$-InAs claddings ($n_r = 3.10$) and a higher refractive index of the SCL[30]. Figures 7 (b) and (c) show optical mode profiles with distributions of confinement factors for both ICLs with hybrid claddings. The optical mode of the GaSb-based ICL shows a 3.8 % higher optical confinement factor in the active region compared to the mode of the InAs-based ICL. A hypothetical InAs-based ICL structure with GaSb SCLs indicates an equal contribution to the increase of the confinement factor coming from GaSb SCLs and n$^+$- InAsSb, as shown in the supplementary material of this work.

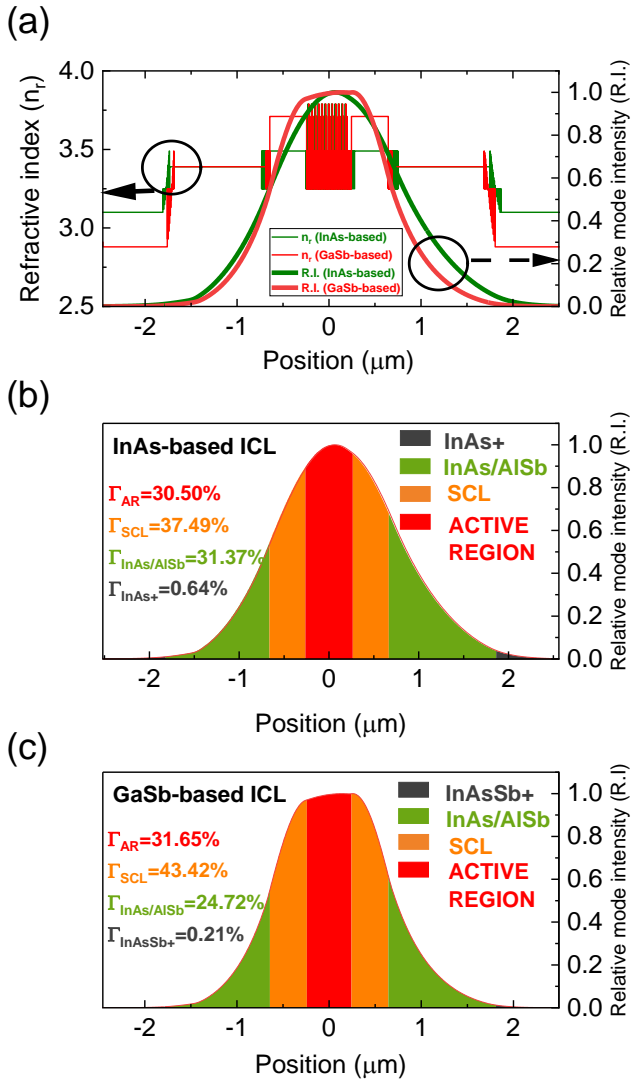

FIG. 7. (a) Refractive index profiles and optical mode intensities along the growth direction of two ICLs with the same active region design but different claddings and SCLs: InAs-based ICL, with hybrid InAs/AlSb – InAs plasmon enhanced claddings and InAs SCLs (green) and GaSb-based ICL, with hybrid InAs/AlSb – InAsSb plasmon enhanced claddings with GaSb SCLs (red) (b) Optical mode confinement distribution along the growth direction of the InAs-based ICL (c) Optical mode confinement distribution along the growth direction of the GaSb-based ICL

By taking effective masses[24], mobilities[23,25] and optical confinement factors into account, the calculated sum of overall FCA loss (FCA x Γ) in claddings and SCLs of the GaSb-based ICL is $\alpha = 3.46$ cm$^{-1}$ which is significantly higher than for InAs-based ICL, $\alpha = 2.69$ cm$^{-1}$. This is a consequence of approximately 9 times higher mobility in InAs compared to GaSb SCLs that led to 3.5 times higher FCA losses in the GaSb SCLs. Confinement in the outer plasmon-enhanced claddings in InAs-based ICL is 3 times higher, nevertheless FCA of n$^+$-InAs is 177 cm$^{-1}$, whereas in n$^+$-InAsSb is 400 cm$^{-1}$. Although mobilities in n$^+$-InAs and n$^+$-InAsSb are almost equal ($\mu \approx 4000$ cm$^2$/Vs), FCA in n$^+$-InAsSb is higher due to lower effective mass. Taking optical confinement to account, this results in the overall FCA in n$^+$-InAsSb claddings being slightly lower than in the n$^+$-InAs claddings. In the SL claddings, the FCA losses are almost equal in both designs. The active region of both ICLs has the same structure and is designed for a laser emission wavelength of 4.6 µm. It consists of 12 standard InAs/Ga$_{0.6}$In$_{0.4}$Sb/InAs W-shaped quantum wells. The bands are in a broken gap alignment. According to enhanced performances realized by mitigating intervalence band absorption in the active region as reported for 4.35 µm[31] and 6.2 µm[32] ICL designs, the suggested thickness of GaInSb well for emission in the range of 4.0-4.8 is 2.50 nm. For this reason, a 2.50 nm thick hole quantum well was implemented.

## III. GROWTH AND FABRICATION

The ICLs were grown by molecular beam epitaxy (MBE). The reactor is equipped with standard effusion cells for the group III elements, silicon and tellurium as n-type dopants and two valved cracker cells for arsenic and antimony. The ICLs were grown on 2" n-type doped substrates. The InAs substrate is doped with Si ($1 \cdot 10^{18} cm^{-3}$) and GaSb substrate is doped with Te ($1 \cdot 10^{18} cm^{-3}$). The substrate temperature during the growth was monitored by a pyrometer. Prior to the growth of InAs-based ICL, an oxide desorption step was carried out under As flux at 540 ºC for 3 minutes. The growth starts with 200 nm InAs buffer, following highly n-type doped 700 nm InAs:Si claddings ($1 \cdot 10^{19} cm^{-3}$). Subsequent growth of 1 µm thick InAs/AlSb inner SL claddings was carried out employing 3 s arsenic soak time after every AlSb layer to force AlAs-like interfaces, in order to balance the compressive strain in AlSb layers on InAs substrate. InAs layers of the inner claddings were n-type doped with Si with a linearly decreasing doping concentration towards the active region ($1 \cdot 10^{18}$ cm$^{-3}$ to $1 \cdot 10^{17}$ cm$^{-3}$) to minimize optical losses. The substrate temperature throughout the whole growth was maintained at 450 ºC. For GaSb-based ICLs, the analog oxide-desorption treatment was performed under Sb flux at a substrate temperature of 560 ºC. The growth starts with 200 nm n-type doped GaSb:Te buffer layer ($1 \cdot 10^{18} cm^{-3}$) at a substrate temperature of 500 ºC.

The temperature was then ramped down to 450 ºC for the growth of 700 nm thick n⁺-InAsSb: Si cladding ($1 \cdot 10^{19}\,cm^{-3}$) and 1 µm thick inner InAs/AlSb SL cladding. InAs layers of the inner claddings were doped in the same fashion as in the InAs-based ICL. The superlattices were grown with no soak times, as the InAs layers are, in this case, tensile strained on GaSb substrate and countering compressively strained AlSb layers. The substrate temperature was kept at 450 ºC for the whole duration of growth, except for the growth the bottom and top GaSb-SCL. Prior to the growth of SCLs, the substrate temperature was ramped up from 450 ºC to 500 ºC and after SCL growth, it was ramped back to 450 ºC. Both separate confinement layers were 400 nm thick to increase the confinement in the active region and lightly doped ($6 \cdot 10^{16}\,cm^{-3}$) also for the sake of lower optical losses. Throughout both ICL structures, InAs/AlSb transition layers had to be implemented between active region, SCLs and cladding layers, for the sake of matching the band offsets. The active WQW is composed of **AlSb**/*InAs*/Ga$_{0.6}$In$_{0.4}$Sb/*InAs*/**AlSb** with thicknesses of (**2.5**/*2.12*/2.50/*1.67*/**1.0**) nm, respectively. The hole and electron injector are composed of 2.8 nm GaSb/1.0 nm AlSb/4.8 nm GaSb and 2.5 nm AlSb/4.4 nm InAs/1.2 nm AlSb/3.2 nm InAs/1.2 nm AlSb/2.5 nm InAs/1.2 nm AlSb/2.05 nm InAs respectively. The last four InAs wells of the electron injectors were n-type doped to $4 \cdot 10^{18}\,cm^{-3}$ for the sake of carrier rebalancing[33]. To reduce the contact resistance, the InAs-based and GaSb-based ICLs were capped with 25 nm thick layer of highly n-type doped ($2 \cdot 10^{19}\,cm^{-3}$) InAs and InAsSb, respectively. From high-resolution X-ray diffraction (HR-XRD) scans of the grown InAs-based and GaSb-based ICLs (not shown), the measured lattice mismatches of inner superlattice claddings are 330 ppm and -185 ppm, respectively indicating good compensation of internal strain. The mismatch of outer claddings of GaSb-based ICL is -560 ppm. Broad area (BA) devices were fabricated for characterization in pulsed mode. For their processing standard photolithography was used to define 100 µm wide ridges. In the wet chemical etching process a mixture of $H_2O/H_3PO_4/H_2O_2/HOC(CO_2H)(CH_2CO_2H)_2$ was used to etch through the active region and the bottom SCL. Afterwards, a Ti/Pt/Au top contact was deposited and AuGe/Ni/Au layers were evaporated at the substrate side as a back contact. The fabricated sample was cleaved into 2 mm long laser bars and measured epilayer side up on a copper heat sink.

## IV. RESULTS

Pulsed measurements were performed for a repetition rate of 1 kHz and the applied current pulse width was 500 ns to minimize Joule heating. Figures 8 (a) and (b) depict electrooptical characteristics of the BA ICLs in pulsed mode operation under different temperatures from 15 ºC to the maximum temperature of $T_{max} = 55$ ºC. At the same current the voltage decreases at higher temperatures due to temperature induced reduction of the band gap, and consequently the overall voltage. At room temperature, the InAs-based ICL exhibits a threshold current density of $J_{th} = 434$ A/cm² whereas the GaSb-based ICL shows a reduced threshold of $J_{th} = 359$ A/cm². The majority of reported values is in this range at similar wavelengths[8,18,34].

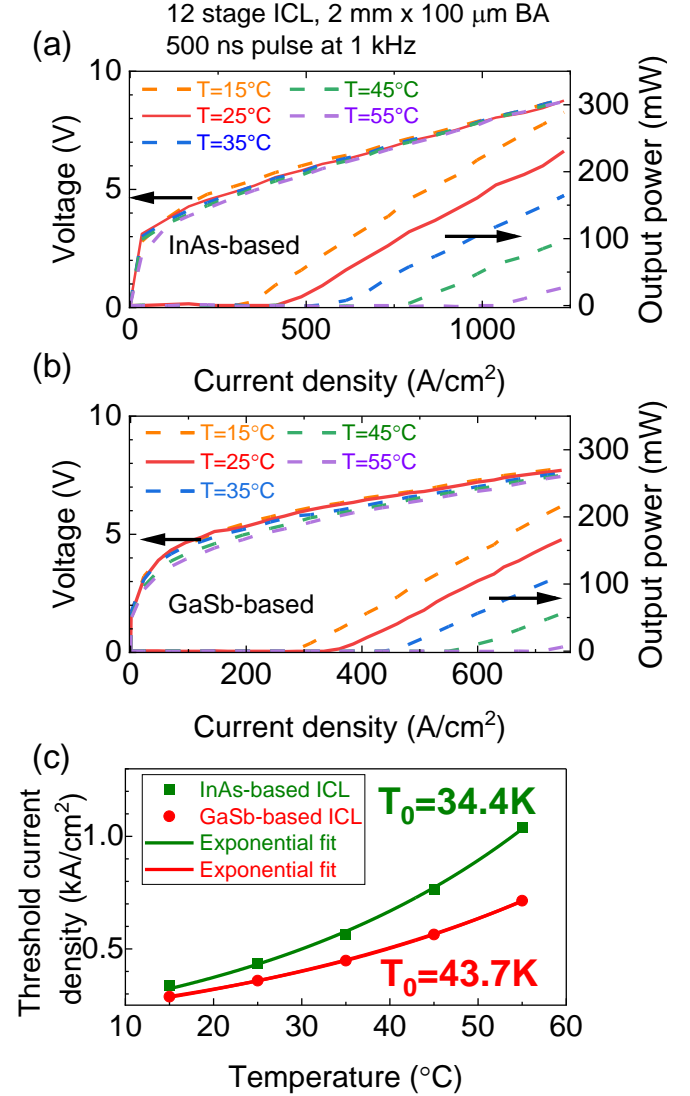

FIG. 8. (a) Pulsed light-current and light-voltage characteristic at different temperatures for 2 mm long and 100 µm wide broad area lasers: (a) InAs-based ICL (b) GaSb-based ICL. (c) Temperature dependences of threshold current densities of InAs-based ICL and GaSb-based ICL and extracted characteristic temperatures $T_0$

The obtained threshold current densities are substantialy higher than recently reported values for ICL with similar wavelengths and hybrid-cladding designs 306 A/cm² (5.17 µm)[35] and 252 A/cm² (4.63 µm)[36]. Reported maximum operating tempratures in pulsed mode are also significantly higher: $T_{max} = 96$ ºC and $T_{max} = 106$ ºC respectively. This suggests there is a room for improvement in terms of a design of the active region and growth conditions. However, reaching record low threshold current densities requires optimal combinations of parameters that might not be reproducible and sustainable over one growth campaign to make a comparison between ICLs grown on different substrates. Although obtaining such low thresholds at similar wavelength was feasible as shown in Ref. 37, it is not crucial for this comparison. The threshold current density of the GaSb-based ICL is 17.3 % lower compared to the InAs-based ICL. This could be attributed to higher optical confinement in the active region. The measured differential resistances of the InAs-based ICL of 2.8 Ω and GaSb-based ICL of 2.3 Ω are slightly higher than the typical 1-2 Ω[11], which might be related to a slightly too low doping of the SCLs. The threshold voltages $V_{th}$ = 6.24 V of the GaSb-based ICL and $V_{th}$ = 5.53 V of the InAs-based ICL are higher than the previously reported values[18,20,35,36]. The higher value for GaSb-ICL than for InAs-based ICL is based on larger conduction band offsets between the SCLs and their adjacent layers[12]. This results in lower voltage efficiency of $\eta_V$ = 49.6 % in comparison to $\eta_V$ = 59.0 % of the InAs-based ICL. The threshold power density at room temperature of the InAs-based ICL of $P_{th}$ = 2.4 kW/cm$^2$ is comparable to the value obtained for the GaSb-based ICL of $P_{th}$ = 2.24 kW/cm$^2$. For the displayed temperature range, for InAs-based ICL, the slope efficiency drops from 167 to 59 mW/A, whereas for GaSb-based ICL it varies from 236 to 84 mW/A. In Figure 8 (c) the temperature dependence of threshold current density of the two ICLs is presented featuring characteristic temperatures of $T_0$ = 34.4 K and $T_0$ = 43.7 K for InAs-based and GaSb-based ICL, respectively. The GaSb-based ICL is, hence, demonstrating higher temperature stability than InAs-based ICL. The values compare well with reported values for similar wavelengths[19,31-32]. The relevant parameters and figures of merit of both ICLs with respective previously reported values[8,9,20,31,36,38] are summarised in Table I.

|  | InAs-based | GaSb-based | Literature |
| --- | --- | --- | --- |
| $J_{th}$ (A/cm$^2$) | 434 | 359 | 247-500 |
| $P_{th}$ (kW/cm$^2$) | 2.40 | 2.24 | 0.8-1.5 |
| $V_{th}$ (V) | 5.53 | 6.24 | 3-6 |
| $n$ (cm$^{-3}$) | $4 \cdot 10^{18}$ | $4 \cdot 10^{18}$ | $3-4 \cdot 10^{18}$ |
| $T_0$ (K) | 34.4 | 43.7 | 35-57 |
| $T_{max}$ ($^0$C) | 55 | 55 | ≤106 |

Table I: Figures of merit of the InAs and GaSb-based ICL and state of the art values at similar wavelengths.

In Figure 9 threshold current denisties are presented for different devices of both InAs and GaSb-based ICLs. The average threshold current density of GaSb-based ICLs is lower by 14.4% than that of InAs-based ICL. This is approximately matching to the relative difference of the respective lowest threshold current densities of the two ICLs.

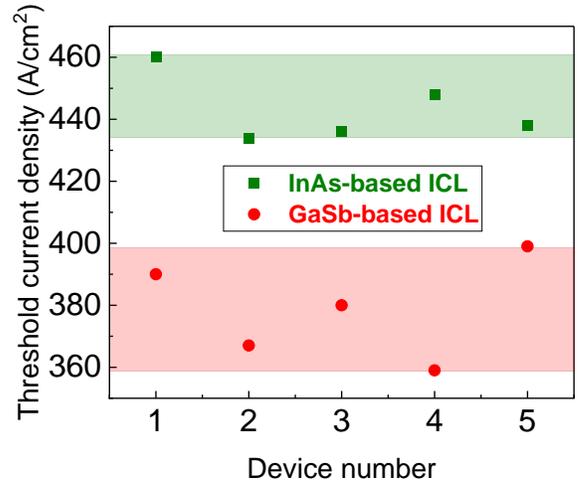

FIG. 9. Threshold current densities at pusled operation of different 2 mm long and 100 µm wide broad area devices: InAs-based ICL (green) and GaSb-based ICL (red).

Figure 10 (a) displays lasing spectra of the InAs-based ICL

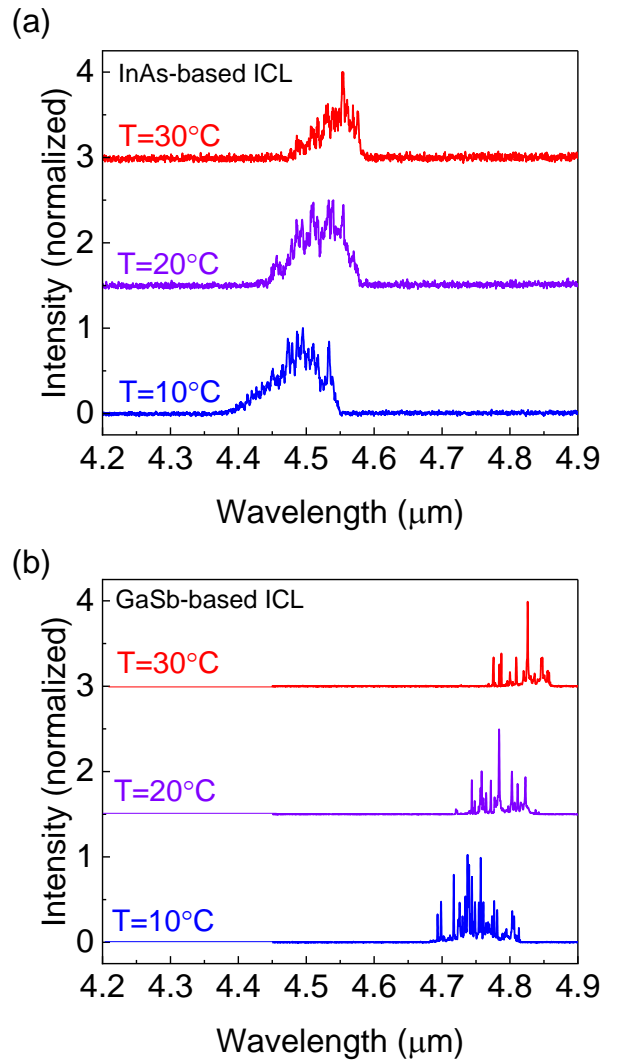

FIG. 10. Spectra at 10⁰C, 20⁰C and 30⁰C of broad area (2 mm x 100 µm) devices: a) InAs-based ICL, the central wavelength at room temperature is 4.55 µm and the temperature shift is 3.95 nm/⁰C. b) GaSb-based ICL, the central wavelength at room temperature is 4.81 µm and the temperature shift is 4.42 nm/⁰C.

at three different temperatures. Figure 10 (b) shows spectra for GaSb-based ICL at the same three temperatures. The central wavelength at room temperature of 4.81 µm obtained for GaSb-based ICL is slightly longer than 4.55 µm for InAs-based ICL This is likely a consequence of smaller band gap of InAs wells in the recombination region introduced by a tensile strain. However, GaSb-based ICL is showing superior performance despite a higher associated FCA loss at the longer wavelength. The corresponding temperature induced shifts are 3.95 and 4.42 nm/⁰C. Spectral linewidth of GaSb-based ICL is approximately 30 nm (≈19% wider) than of InAs-based ICL, which suggests lower losses.

## V. CONCLUSIONS

In summary, we have compared InAs-based and GaSb-based ICLs with analog hybrid cladding architecture for wavelengths $\lambda \approx 4.6$ µm both theoretically and experimentally. Due to lower refractive index of outer plasmon-enhanced claddings and higher refractive index of SCLs, GaSb-based ICL shows a twofold improvement of the optical confinement. Experimentally, although both ICLs have shown higher threshold current densities than previously reported, a comparison between the two devices could be made. The GaSb-based ICL has shown a lower threshold current density and a higher characteristic temperature. For GaSb-based ICL, analysis of the optimization of the cladding layer structure was shown. Additional improvements can be made by further optimization of the active region wells, injector doping and number of cascade stages.

## ACKNOWLEDGEMENTS

We are grateful to European Union's Horizon 2020 research and innovation programme under the Marie Skłodowska-Curie grant agreement no 956548 (QUANTIMONY) for financial support of this work.

## AUTHOR DECLARATIONS

**Conflict of interest**

The authors declare no conflicts to disclose.

## DATA AVAILABILITY

The data that support the findings of the study are available from the corresponding author upon reasonable request.